\documentclass[aps,prb,citeautoscript,reprint,showpacs,superscriptaddress,twocolumn]{revtex4-2}
\usepackage{graphicx}
\usepackage{preamble}

\usepackage{subfigure}
\def\R{{{\rm I}\kern-.2em {\rm R}}}
\usepackage{float}
\begin{document}

\date{\today}

\title{Coherent Magneto-Conductance Oscillations in Amorphous Topological Insulator Nanowires}

\author{Siddhant Mal}
\affiliation{Department of Physics, University of Michigan, Ann Arbor, Michigan 48104, USA}
\affiliation{Department of Physics, University of California, Berkeley, California 94720, USA}
\author{Elizabeth Dresselhaus}
\affiliation{Department of Physics, University of California, Berkeley, California 94720, USA}
\author{Joel E. Moore}
\affiliation{Department of Physics, University of California, Berkeley, California 94720, USA}
\affiliation{Materials Sciences Division, Lawrence Berkeley National Laboratory, Berkeley, California 94720, USA}

\begin{abstract}
Recent experiments on amorphous materials have established the existence of surface states similar to those of crystalline three-dimensional topological insulators (TIs) (\cite{corbae2019evidence, corbae2023amorphous,checkelsky2009quantum}). Amorphous topological insulators are also independently of interest for thermo-electric and other properties. To develop an understanding of transport in these systems, we carry out quantum transport calculations for a tight-binding model of an amorphous nano-wire pierced by an axial magnetic flux, then compare the results to known features in the case of crystalline models with disorder. Our calculations complement previous studies in the crystalline case that studied the surface or used a Green’s function method (\cite{zhang2010anomalous,bardarson2010aharonov}).  We find that the periodicity of the conductance signal with varying magnetic flux is comparable to the crystalline case, with maxima occurring at odd multiples of magnetic flux quanta. However, the expected amplitude of the oscillation decreases with increasing amorphousness, as defined and described in the main text. We characterize this deviation from the crystalline case by taking ensemble averages of the conductance signatures for various wires with measurements simulated at finite temperatures. This striking transport phenomenon offers a metric to characterize amorphous TIs and stimulate further experiments on this class of materials.
\end{abstract}

\maketitle

\section{Introduction}\label{sec:intro}

Topological insulators (TIs) induced by spin-orbit coupling are of significant applied interest due to their potential utility in spintronics and novel quantum device applications \cite{tian2017property,hasan2010colloquium}. It is well known that TI surface states are robust in principle to the breaking of translational invariance through some amount of non-magnetic local disorder, including amorphous structure \cite{kitaev2009periodic,agarwala2017,marsal2020topological}. Recent experimental progress has demonstrated that an amorphous form of bismuth selenide (Bi$_2$Se$_3$), which is a well-studied TI in crystalline form, can have surface states that are visible in angle-resolved photoemission spectroscopy and appear similar to those in the crystalline case.\cite{corbae2019evidence,ciocys2024establishing}

It is natural to ask whether other distinctive features that distinguish the topological insulator state from an ordinary insulator are present in the amorphous case. Aside from fundamental interest, this is of applied importance as amorphous materials can have practical advantages over crystalline ones, such as ready integration with conventional semiconductors. 

Here we are interested in characterizing a distinct transport phenomenon in an amorphous three-dimensional TI invariant under time-reversal ($\mathcal{T}$) symmetry. Such a state is characterized by a gapped bulk and a protected gapless surface state. The transport phenomenon we now explain has previously been studied in crystalline TIs\cite{bardarson2010aharonov,zhang2010anomalous,yicui,yongchen,masonnanowire,andonanowire} and is a distinct, quantitative signature of the topological state.

The surface state is modelled by a gapless Dirac fermion. However, if the system assumes the geometry of an infinite cylindrical wire then this mode is no longer protected and a gap emerges. This occurs due to the induced compatible spin structure on the cylindrical surface which encodes a $\pi$ Berry flux and sets the fermion in the Ramond sector, that is, with periodic boundary conditions \cite{wernli2019lecture,zhang2010anomalous,lee2009surface}. This implies that the fermionic modes occur in degenerate pairs while the $\mathcal{T}$ symmetry protects only the parity of the number of surface modes. With these boundary conditions the modes are gapped. The protected gapless mode can be re-established by threading an additional $\pi$ Berry flux through the bulk of the cylinder via a magnetic flux of magnitude $h/2e$. Such a wire thus exhibits conductance oscillations with period $h/e$, but with a half-period shift (a $\pi$ phase shift) compared to conventional Aharonov-Bohm oscillations. This is also to be contrasted with the Sharvin-Sharvin \cite{PismaZhETF.34.285} type weak localization oscillations of period $h/2e$ \cite{webb1985observation}.

\begin{figure}[!h]
    \centering
    \subfigure[]{\includegraphics[width=0.12\textwidth]{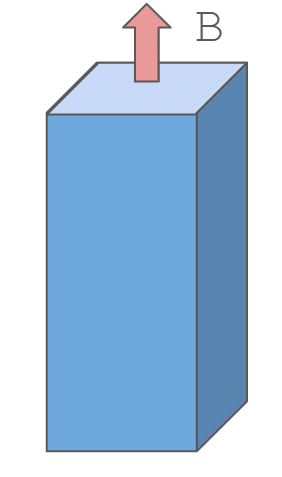}}
    \subfigure[]{\includegraphics[width=0.21\textwidth]{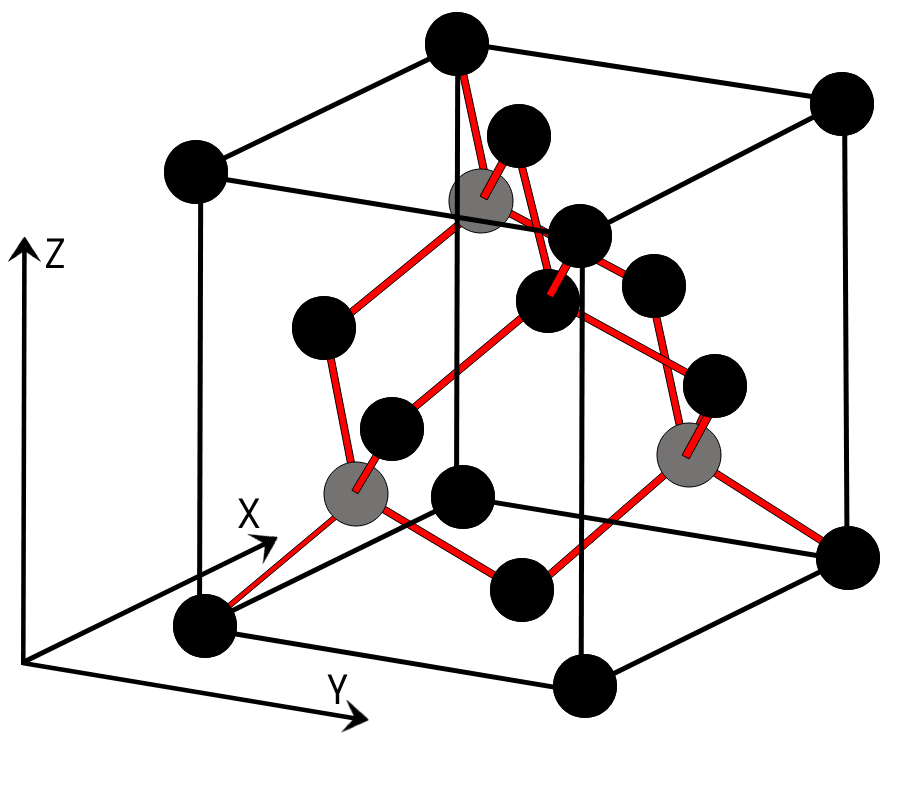}}
    
    \caption{(a) Cuboidal geometry of nano-wire with magnetic flux penetrating the bulk in the length of the wire. (b) The cubic unit cell for diamond used to assemble a crystalline wire. }
    \label{fig:geom}
\end{figure}
\begin{figure}[!h]
    \centering
    \hspace{-0.5cm}
    \subfigure{\includegraphics[width=0.4\textwidth]{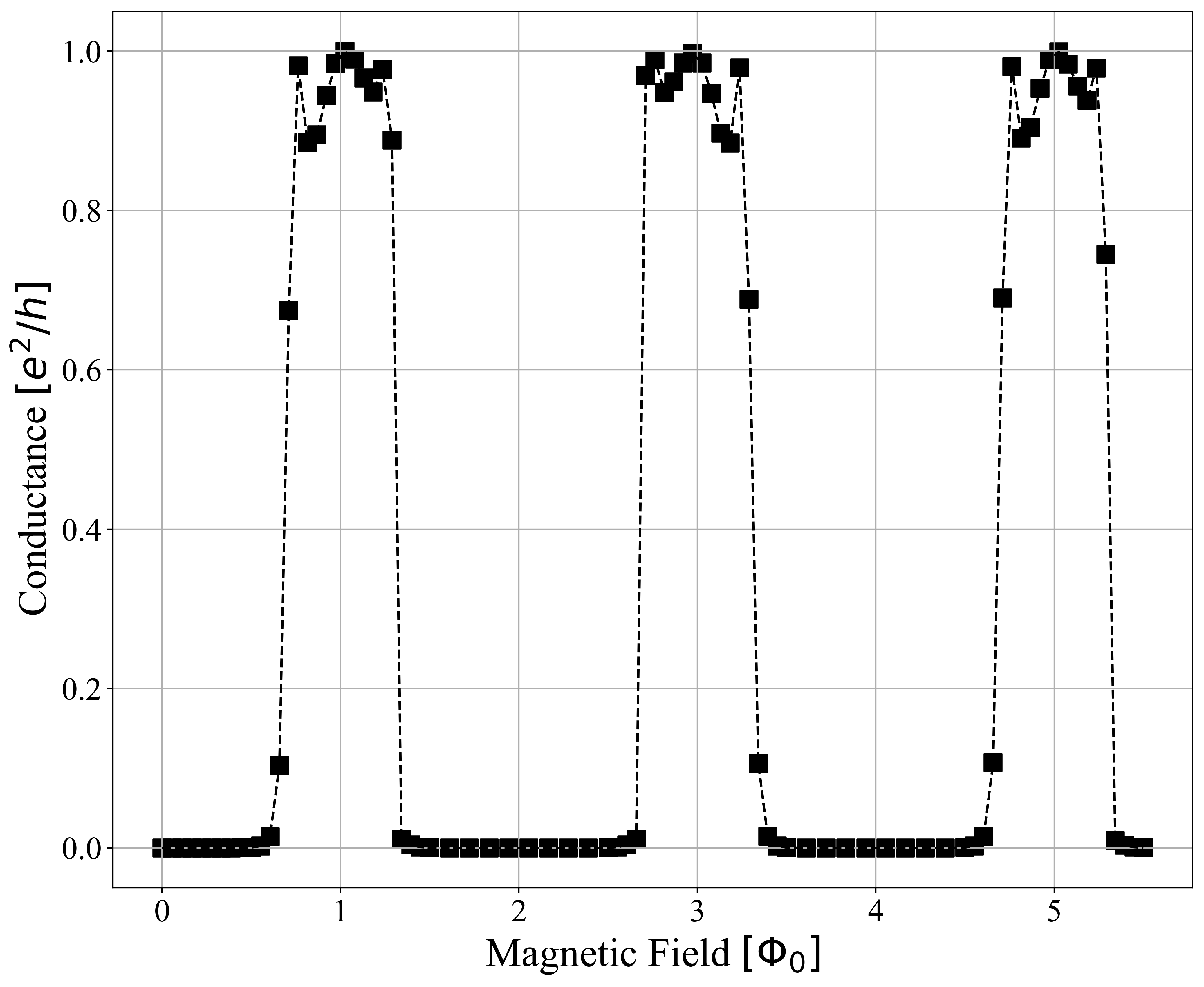}}
    
    \caption{Oscillations of the conductance of the wire with a Fu-Kane-Mele type crystalline Hamiltonian for a sample of $5\times5\times40$ unit cells with varying flux quanta.}
    \label{fig:crys_signal}
\end{figure}

In \cite{bardarson2010aharonov}, this effect was investigated by considering the effective theory of a $2$D Dirac fermion in the presence of a Gaussian surface disorder potential. In Ref \cite{zhang2010anomalous} a three-dimensional simulation of the effect was done with onsite surface disorder which also shows robustness of the conductance oscillations. The analysis there was done using a recursive Green's function method. The conductance signal for a crystalline wire is reproduced in Figure \ref{fig:crys_signal}.

Experimentally, the predicted topological version of Aharonov-Bohm oscillations has been observed in topological insulator nanowires \cite{yicui,yongchen,masonnanowire,andonanowire} and provided a practical way to isolate effects of the topological surface states even when bulk conduction is nonzero, as is the case for most or all 3D TI materials as-grown. Since the bulk of the wire does not respond with the same period to the applied flux (indeed, as a continuous range of periods are present, the bulk contribution does not show regular oscillatory behavior), the oscillations are determined by surface state conduction. Among other generalizations, there is a superconducting version of the $\pi$-flux effect when the TI nanowire is proximitized by superconducting contacts.\cite{fernandoprl2014,bardarsonreview}

Here we investigate this effect for a 3D amorphous system using a scattering transport calculation. We use the geometry of a cuboid which is topologically equivalent to a cylinder as shown in Figure \ref{fig:geom}. We will start from the Fu-Kane-Mele \cite{fu2007topological} model for the 3D crystalline TI. Our method is to introduce connectivity defects into the lattice to `amorphize' it.

The amorphization procedure we employ locally preserves coordination number. We consider tight-binding models on these structures that have an exponentially decaying hopping amplitude, mimicking the dispersion of inter-atomic bonds. Models in this class conserve local order in the absence of long-range order. This is in contrast with the first models to predict topological phases in amorphous electronic systems, which do not conserve coordination number and construct hoppings between randomly arranged sites using the vector that connects them \cite{agarwala2017}. We choose to focus on models of the former type because they are more realistic \cite{marsal2020topological} and have been invoked to find exact results on $2D$ amorphous systems. Results include proof of the persistence of spectral gaps in models of amorphous silicon \cite{wearie, wearie-thorpe}, prediction of a topological phase diagram for amorphous topological insulators \cite{marsal2020topological}, and the manifestation of flat band-like degeneracies in amorphous models of a platform relevant to topological photonics experiments \cite{dresselhaus-amorphous-photonics-localization}. In this work, we numerically investigate a $3D$ variation on these models hosted on an amorphized diamond structure.

We first demonstrate the formation of the Dirac cone for an amorphous system by treating it as a super-cell, that is, imposing periodic boundary conditions in the long direction of the wire (see Figure \ref{fig:periodic}) (henceforth ${\bf \hat z})$  and varying magnetic field. Then we use the Landauer formalism using the python package \textit{kwant} \cite{groth2014kwant, imry1999conductance, landauer1957spatial} to find and characterize the conductance signatures of wires of these systems for different connectivity defect densities (as defined in the next section), sizes, and temperatures.

\section{Model and Lattice}

\begin{figure}[!h]
    \centering
    \subfigure{\includegraphics[width=0.18\textwidth]{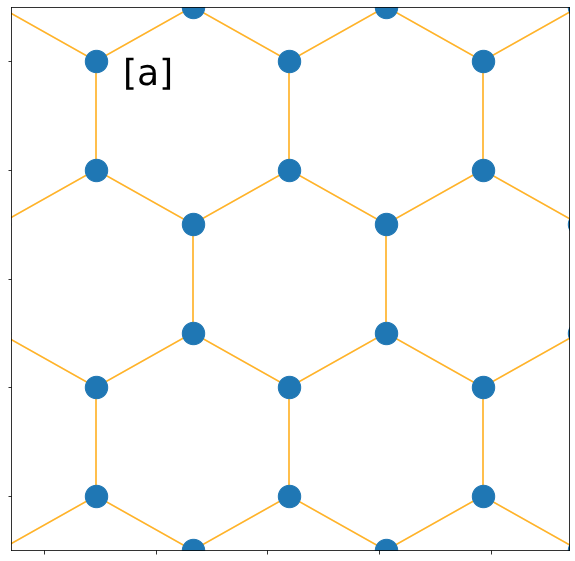}}
    \subfigure{\includegraphics[width=0.18\textwidth]{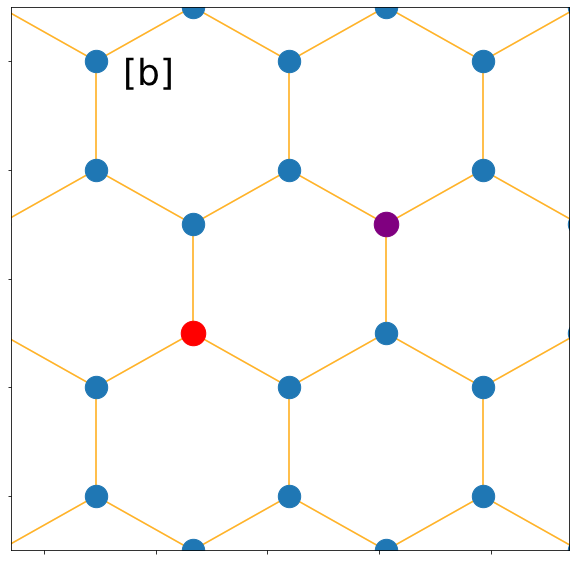}}
    \subfigure{\includegraphics[width=0.18\textwidth]{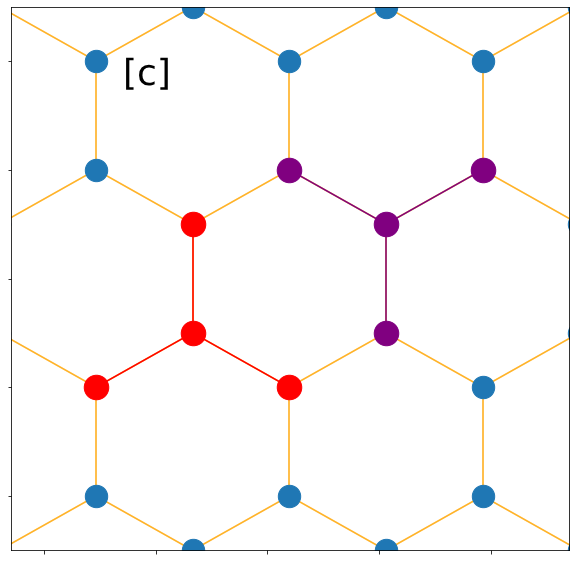}}
    \subfigure{\includegraphics[width=0.18\textwidth]{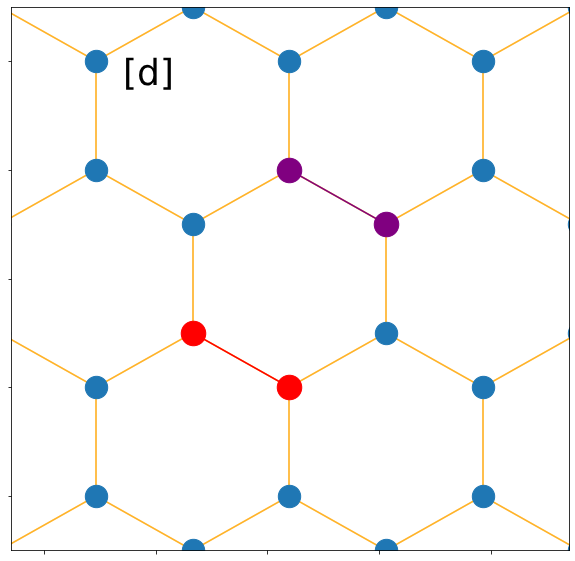}}
    \subfigure{\includegraphics[width=0.18\textwidth]{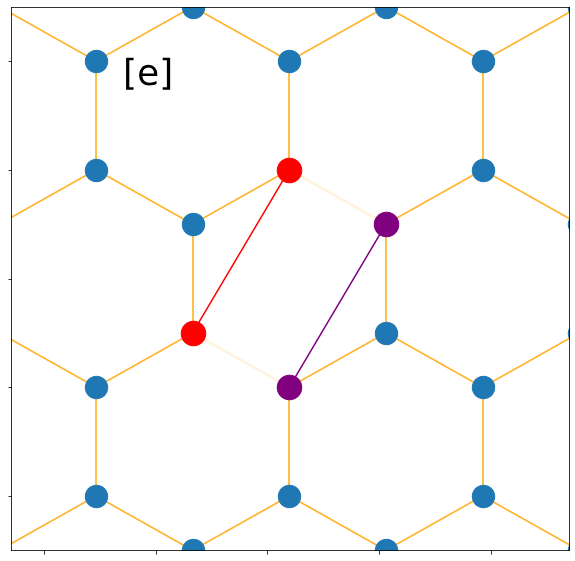}}
    \subfigure{\includegraphics[width=0.18\textwidth]{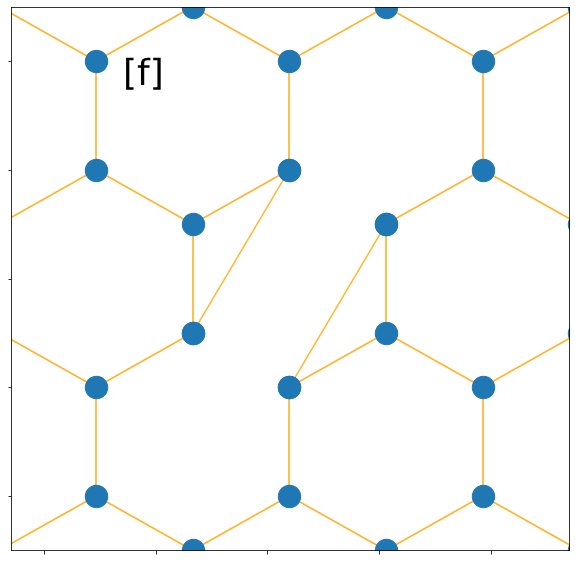}}
    \caption{Illustration of the `amorphization' procedure in a 2D setting. (a) The original lattice. (b) The selection of a random site (red) and another (purple) within (in this case) 3 nearest neighbor bonds of it. (c)-(d) Choosing from the nearest neighbour bonds of each site, one, at random. (e) Making the crossed replacement with the described probability. (f) Repeat until the desired defect density is reached.}
    \label{fig:amorph}
\end{figure}

\textit{Amorphous Fu-Kane-Mele Diamond Model}: To construct an amorphous analog of the Fu-Kane-Mele model we start with a diamond lattice. Then we make local moves, which have a certain probability to introduce non-optimal (in the sense of bond length) bonds while always keeping a fixed co-ordination number $z = 4$ for each site. Specifically, for a randomly chosen site $a$  we pick a site $b$ in its vicinity and select, at random, a nearest neighbor bond for each of these sites. Here, the sites in the vicinity of $a$ are defined in our case to be those that can be traversed from $a$ in fewer than three nearest neighbour bonds. Then we consider crossing the connections in these two bonds with a probability $e^{-\Delta}$, where, 
\begin{equation}
    \Delta=\beta(D_f-D_i),\;\;\;(\beta>0)
\end{equation}
and $D_i,D_f$ are the sum of the bond lengths, before and after the swap respectively. We only consider swaps where $D_f>D_i$ as to only introduce non-optimal bonds. This procedure can be considered the inverse of the one described in Ref \cite{schrauth2019fast}. We note, as an aside, that the probability of picking an already modified bond and re-modifying it is vanishing for the connectivity defect densities that we consider in this paper. Figure \ref{fig:amorph} illustrates the procedure.
\\

This procedure allows us to smoothly control the \textit{degree of amorphousness}. On the modified lattice we implement the following Hamiltonian which is the Fu-Kane-Mele Hamiltonian (\cite{fu2007topological}) with a special choice of the nearest neighbour terms, mimicking an applied external stress (\cite{fu2007topological}), to suit the amorphous structure,
\begin{equation}
    \hat{H}=\sum_{\braket{ij}}e^{\Delta_1}c^\dag_i(1+\Vec{d_1}\cdot\Vec{s})c_j+i8\lambda\sum_{\braket{\braket{ij}}}e^{\Delta_2}c^\dag_i[(\vec{d_1}\times\vec{d_2})\cdot\vec{\sigma}]c_j,
\end{equation}
where $\vec{s}=\frac{1}{\sqrt{3}}(1,1,1)$ is the strong hopping direction, $\Delta_1=2\left(1-\frac{|\vec{d}_1|}{a_1}\right)$, $\Delta_1=2\left(1-\frac{|\vec{d}_1+\vec{d}_2|}{a_2}\right)$. Here $a_1,\;a_2$ are the lengths of the nearest and second nearest neighbour bonds for a crystalline lattice. This represents a specific choice for the free parameters in the Fu-Kane-Mele model, this choice is generic and variations on it do not affect the results of the paper. 

This model captures amorphous deviations from crystalline behaviour in the sense that the cycle-length distribution of the graph that describes this tight binding model is altered and bonds of modified length and hopping amplitudes are introduced. This may appropriately be thought of as introducing certain densities of connectivity defects in the lattice. Thus, when characterizing instances of this model we will use the \textit{density of connectivity defects} $\Sigma$ to denote the percentage of nonoptimal bonds (or connectivity defects) per site of the lattice. That is, the ratio of the total number of defects to the number of sites in the lattice. 

\textit{Magnetic Field Implementation:} We implement the magnetic field as a minimal coupling through the Peierls substitution \cite{li2020electromagnetic}. That is, for two sites at $\vec{r}_1,\;\vec{r}_2$, the matrix element $t=\braket{\vec{r}_1|\hat{H}(\vec{A}=0)|\vec{r}_2}$ is modified to:
\begin{equation}
    \braket{\vec{r}_1|\hat{H}(\vec{A})|\vec{r}_2}=te^{-i\int_{\vec{r}_1}^{\vec{r}_2}\vec{A}\cdot d\vec{r}}
\end{equation}
In our case, suppose the sample is contained in the region of space $[-X,X]\times[-Y,Y]\times[0,Z]$, then the choice of electromagnetic gauge field $\vec{A}$ is:
\begin{equation}
    \vec{A}=\Phi\frac{\Phi_0}{2\pi r}\hat{\theta}
\end{equation}
where $\Phi$ is the specified amount of flux in units of $\Phi_0$.\\

\section{Amorphous Super-Cell}
\begin{figure}[!h]
    \centering
    \vspace{0.3cm}\hspace{0.7cm}
    \includegraphics[width=0.43\textwidth]{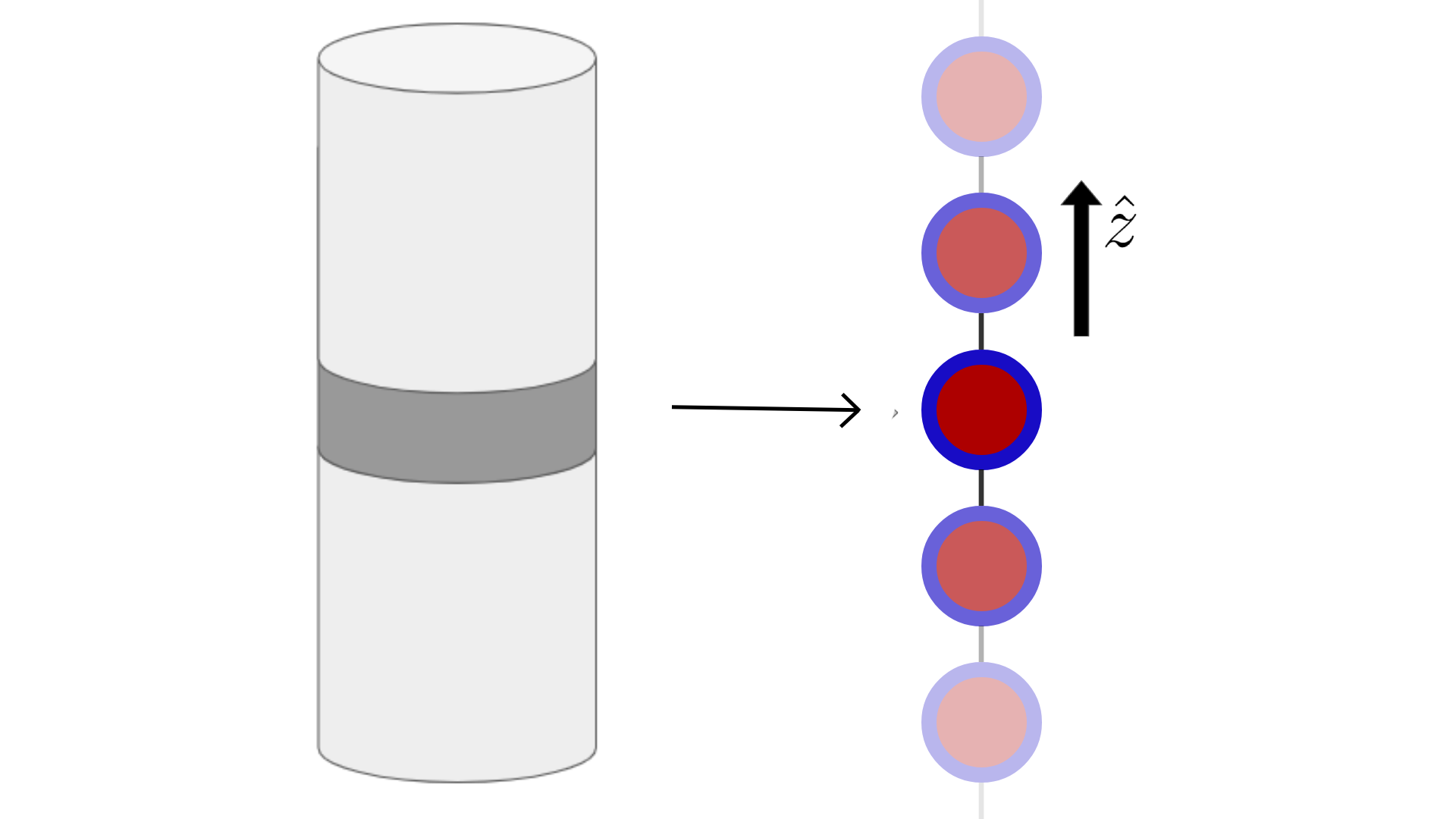}
    \caption{A section of an amorphous wire (shaded dark gray on the left) is treated as the unit cell (red-blue circle on the right) for a 1D lattice in the $\hat{z}$ direction.}
    \label{fig:periodic}
\end{figure}

We can gather our first piece of evidence that the surface mode is present and responds to the magnetic field in the expected manner by using a \textit{super-cell} method as described in \cite{kitaev2009periodic}. We create a cuboidal section of wire along its length and regard it as a unit cell for a 1D lattice (in the direction of the long axis, as shown in Fig \ref{fig:periodic}) and display the spectrum near the Fermi level to observe the formation and disappearance of a surface Dirac cone.

Supposing that we have made a cubical lattice with the given amorphous super-cell as unit cell. Then we recall that the matrix element for the periodic Hamiltonian with momentum $\vec{k}$ is given by:
\begin{equation}
    \braket{i|\hat{H}(\vec{k})|j}=\sum_{k
    \;:
    \;\vec{r}_k-\vec{r}_i\in\Lambda}e^{-i\vec{k}\cdot(\vec{r}_k-\vec{r}_{j})}\braket{k|\hat{H}|j}
\end{equation}
from Bloch's theorem, where $\Lambda$ is the lattice. For an instance of the amorphized wire we see the formation and disappearance of the Dirac cone with magnetic field in Figure \ref{fig:super}. 

It is to be noted that the crystalline Fu-Kane-Mele filling $\mathbb{R}^3$ is guaranteed a doubly degenerate spectrum symmetric about zero. This is due to the combination of time reversal and chiral symmetries. Since we break the chiral symmetry and introduce disorder in the Hamiltonian parameters, the spectrum remains neither symmetric nor degenerate. These changes also move, in energy, the gap of this Hamiltonian. We keep track of the location of the gap by sorting the eigenvalues of the periodic system (with zero magnetic field) ascendingly and inspecting the energy difference in the two middle eigenvalues at $k_z=\pi$ crystal momentum.

\begin{widetext}
\textcolor{white}{.}
\begin{figure*}
    \centering
    \includegraphics[width=0.95\textwidth]{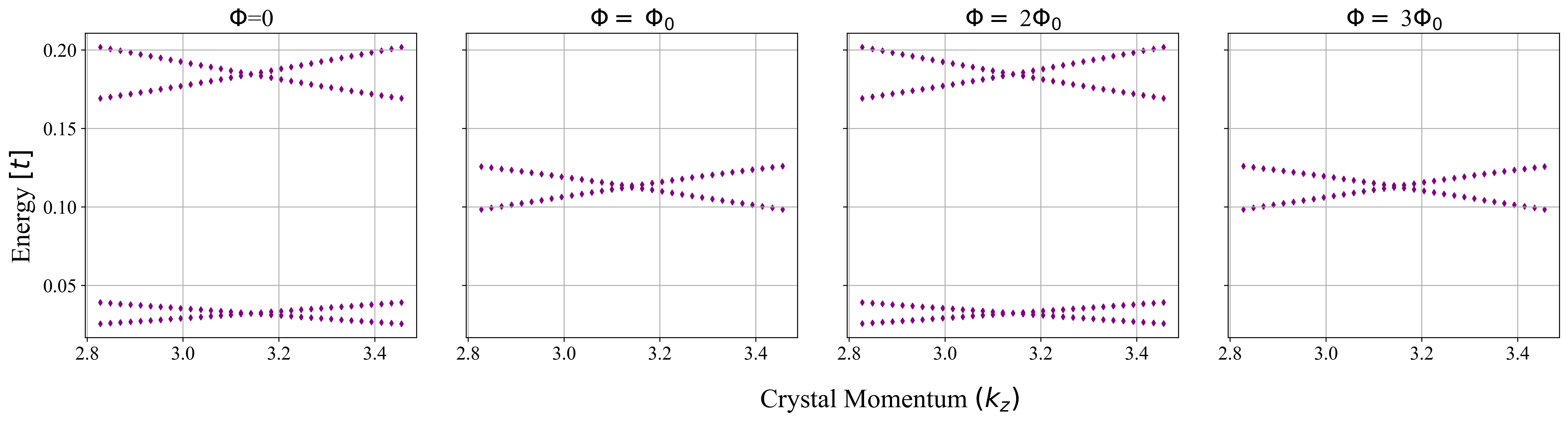}
    \caption{Band spectra for a 1D lattice (regarded as laying in the $\hat{z}$ direction) made from a $7\times7\times3$ super-cell with $\Sigma=3.9\%$. We display the spectrum around $k_z=\pi$ about the energy gap with $Z$ bulk magnetic flux (a) 0, (b) $\Phi_0$, (c) $2\Phi_0$, (d) $3\Phi_0$ (where $\Phi_0$ is the flux quanta $h/2e$) We find that this spectrum repeats with period $2\Phi_0$ as expected. Further we note that both the symmetry in the spectrum about $E=0$ and in momentum about $k_z=\pi$ are lost due to the introduction of defects.}
    \label{fig:super}
\end{figure*}
\end{widetext}

\section{Scattering Transport Signatures}
We now consider the transport signatures of wires of several sizes with varying defect density. We notice that for low densities $(\Sigma<\;\sim15\%),$ and generic values of energy in the gap, the conductance signature preserves its features. For instance we produce an instance of the signal for a $5\times 5\times 50$ wire with $\Sigma=14\%$,
\begin{figure}[!h]
    \includegraphics[width=0.4\textwidth]{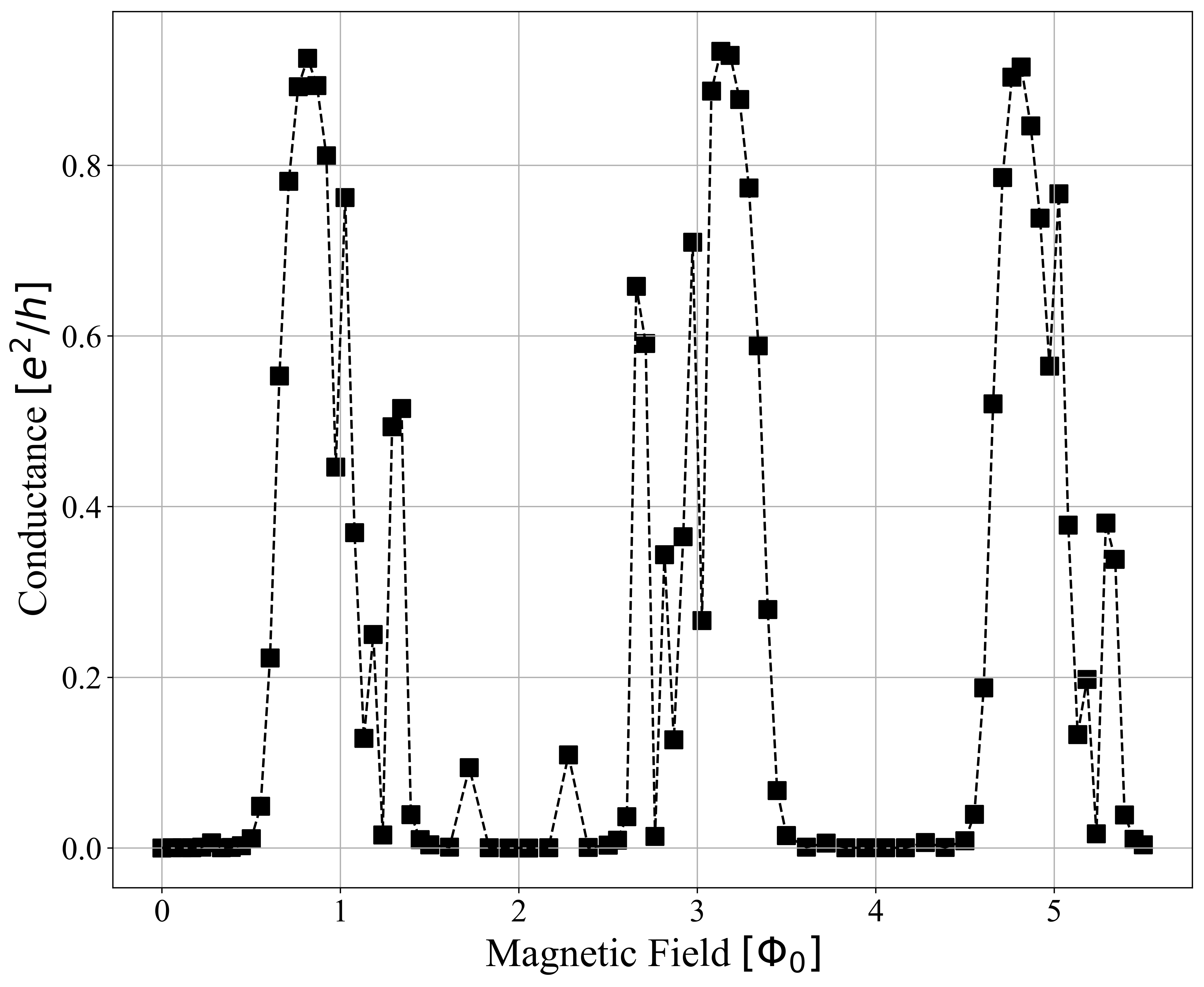}
    \caption{Conductance signature for a $5\times 5\times 50$ cubic unit cell wire with $\Sigma=14\%$ and transmission energy near the middle of the gap.} 
    \label{fig:trans_single}
\end{figure}
Here we notice that while the general form of the signal is preserved, new features arise in the signal. There are more spikes and displacements of the peaks. For a given instance of a wire with a concentration of connectivity defects, the conductance signals at different energies in the gap display different amounts of deviation from the ideal signal. These mesoscopic fluctuations are physical, in that they will indeed exist in a specific disorder realization at zero temperature and vanishing voltage bias. However, because the energy scale of such jagged conductance fluctuations becomes very small as wire size increases, experiments such as Ref.\onlinecite{yicui,yongchen,masonnanowire,andonanowire} tend to see smoother dependence of conductance on gate voltage, because either temperature or voltage bias acts to round out the jaggedness.

This warrants the question of how the system behaves when we are at a finite temperature, such that transmission probabilities at energies in the gap near the probe voltage also contribute to the conductance. For this we model the system as a quantum dot with one dimensional metallic leads connected on opposite sides with potential $V+\Delta V,V-\Delta V$. Suppose that we know the transmission probability distribution $\tau(E)$, then if the whole system is at a temperature $T,$ the equilibrium current of this one dimensional structure is given by,
\begin{equation}
    \begin{split}
        J&=\int_{-\infty}^\infty dE\;\tau(E)\frac{-e}{h}\left[f(E^+)-f(E^-)\right]\\
        &\simeq (-2e\Delta V)\int_{-\infty}^\infty dE\;\tau(E)\frac{-e}{h}\left[\frac{e^{(E+eV)/k_BT}}{k_BTf(E+eV)^2}\right]\\\text{\textcolor{white}{.}}
    \end{split}
\end{equation}

\begin{widetext}
\textcolor{white}{.}
    \begin{figure*}[!h]
    \centering
    \includegraphics[width=0.88\textwidth]{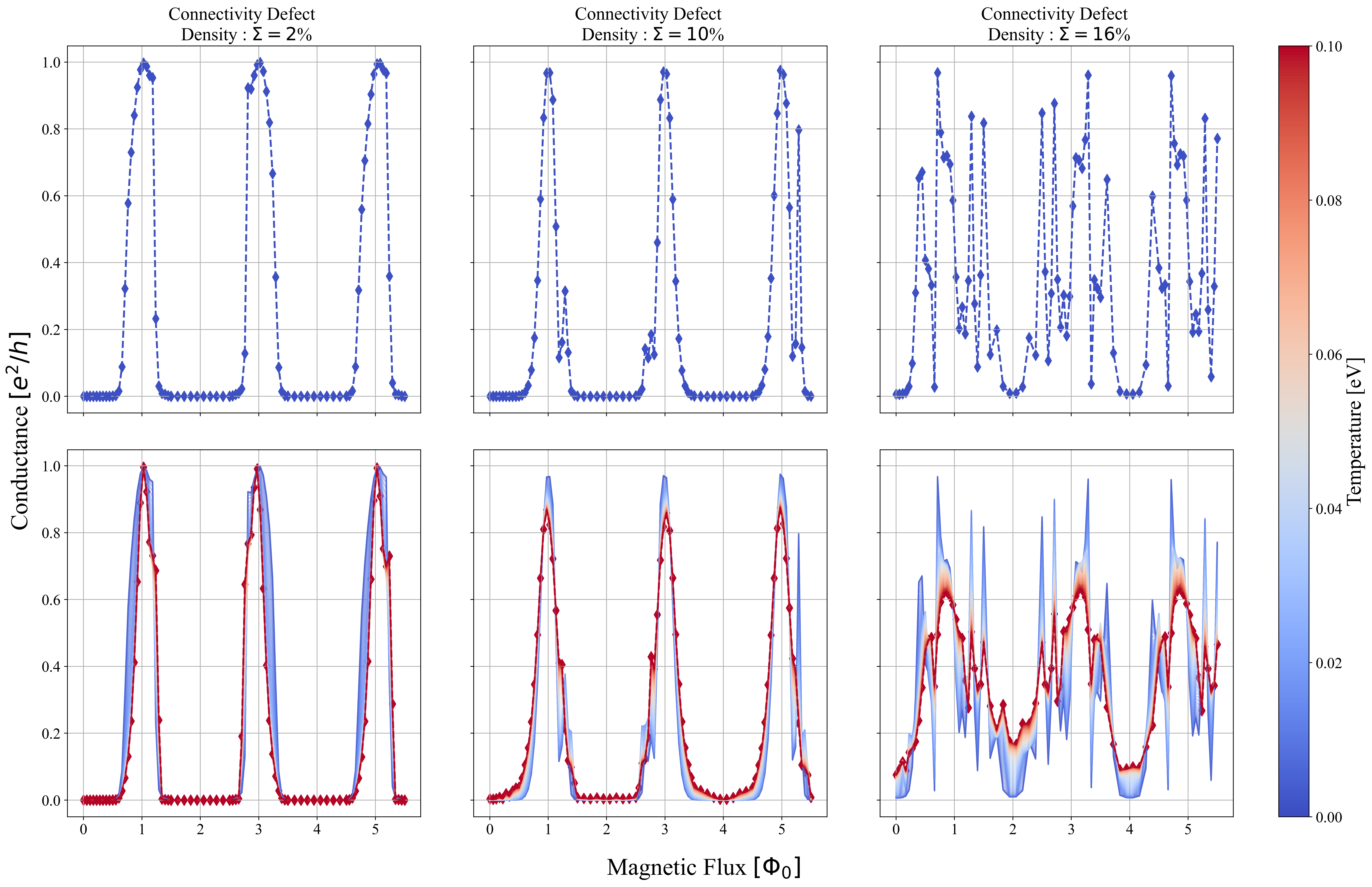}
    \caption{Conductance signatures at zero temperature for a $5\times 5\times 40$ cubic unit cell wire at defect densities $\Sigma=2\%,\;10\%,\;16\%$. For a given defect density $\Sigma,$ the signature on the top row is from a probe energy chosen at the middle of the gap at zero temperature and the plot below it is of the same wire with the same probe energy but now showing the signature as a function of temperature. Warmer colors indicate higher temperature.} 
    \label{fig:trans_single}
\end{figure*}
\begin{figure*}[!h]
    \centering
    \includegraphics[width=0.88\textwidth]{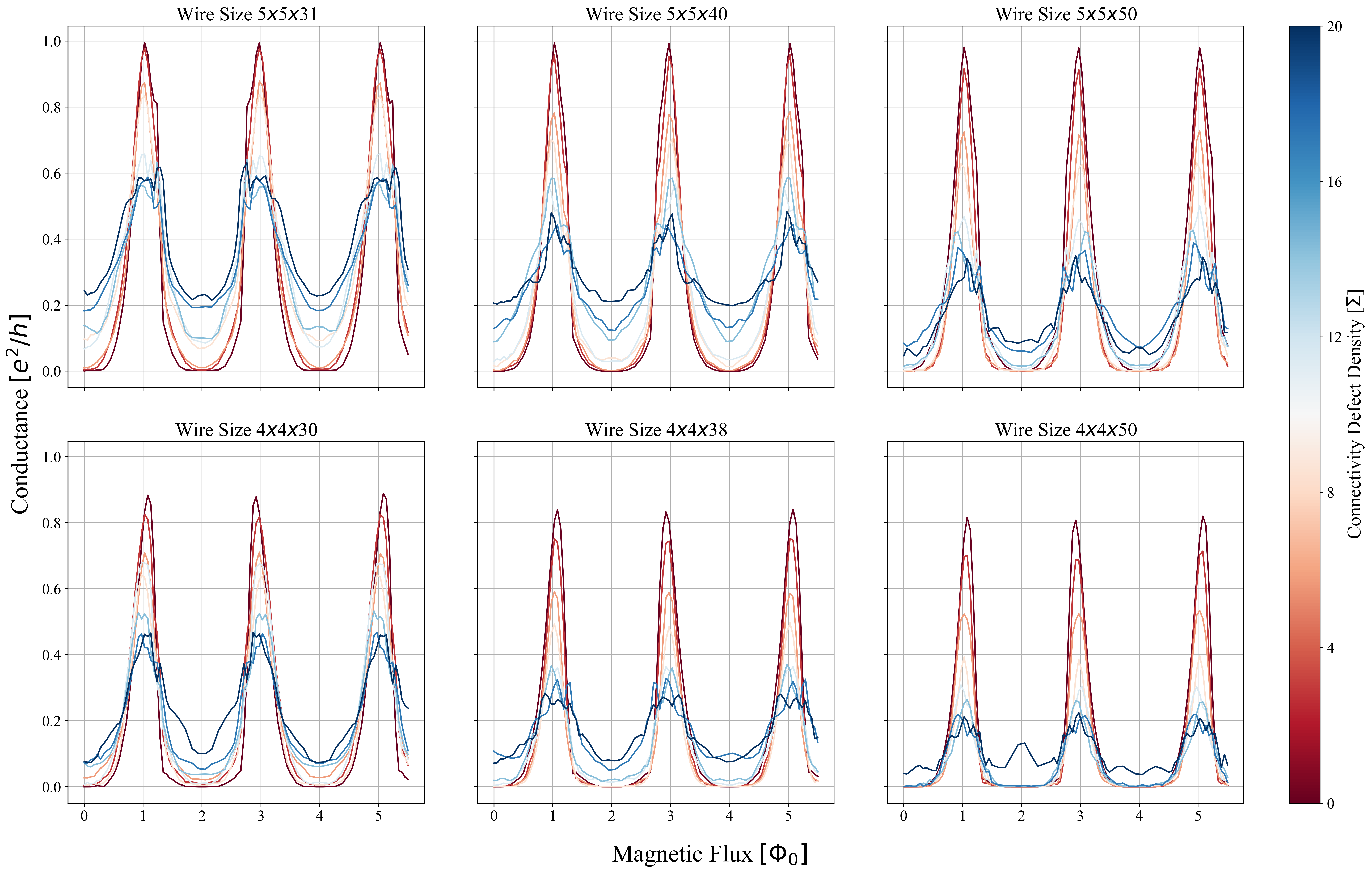}
    \caption{Conductance signatures for high temperature ensemble averages of various wire sizes with connectivity defects densities ranging from $0\%-18\%$. Each ensemble consisted of 7-10 wires and for each temperature average we compute 21 points centered about the middle of the gap and spanning $60\%$ of the gap. Note that the time-reversal symmetry at unit flux quanta is not exact and only becomes so in the thermodynamic limit, thus for wires with a cross section $4\times 4$ cubic unit cells or smaller, the peak conductance will not fully reach $[e^2/h]$ even in the crystalline case, this is explained in \cite{zhang2010anomalous}}
    \label{fig:trans}
\end{figure*}
\end{widetext}

where the Fermi-Dirac function $f(E)=\frac{1}{e^{E/k_BT}+1}$ $E^+=E+e(V+\Delta V),\;E^-=E+e(V-\Delta V)$. To approximate this integral we choose to compute the conductance signature for $n$ evenly spaced points about the center of the gap and treat the integrand as a weight for averaging them. Using this procedure we can probe the transport signal at various temperatures for varying connectivity defect densities. In Figure \ref{fig:trans_single} we  produce an example which is typical of the situation. We show three wires of dimension $5\times 5\times 40$ and show the conductance signal at zero temperature as well as at high temperature for connectivity defect densities $\Sigma=2\%,\;10\%,\;16\%,$. For $\Sigma=16\%$ we pick a particularly jagged signal and show that when we consider the system at sufficiently high temperature, the expected signal for the sample recovers the expected maxima and minima, as shown in Figure \ref{fig:trans_single}. The small size of the sample means that the jagged features have a relatively large energy scale and a high temperature (hundreds of Kelvin) is required for smearing; with a larger sample as measured in experiments, the scale of the jaggedness decreases, and a smaller temperature (or alternately a small applied voltage bias) will be sufficient to yield the smooth behavior seen in most experiments. This phenomenon is consistent over amorphous disorder realizations and defect densities up till $\Sigma=20\%$, where-after the amorphous disorder is strong enough that in many instances the gap closes and we lose the system of interest. 

\begin{figure}[!h]
    \centering
    \hspace{-0.5cm}
    \includegraphics[width=0.4\textwidth]{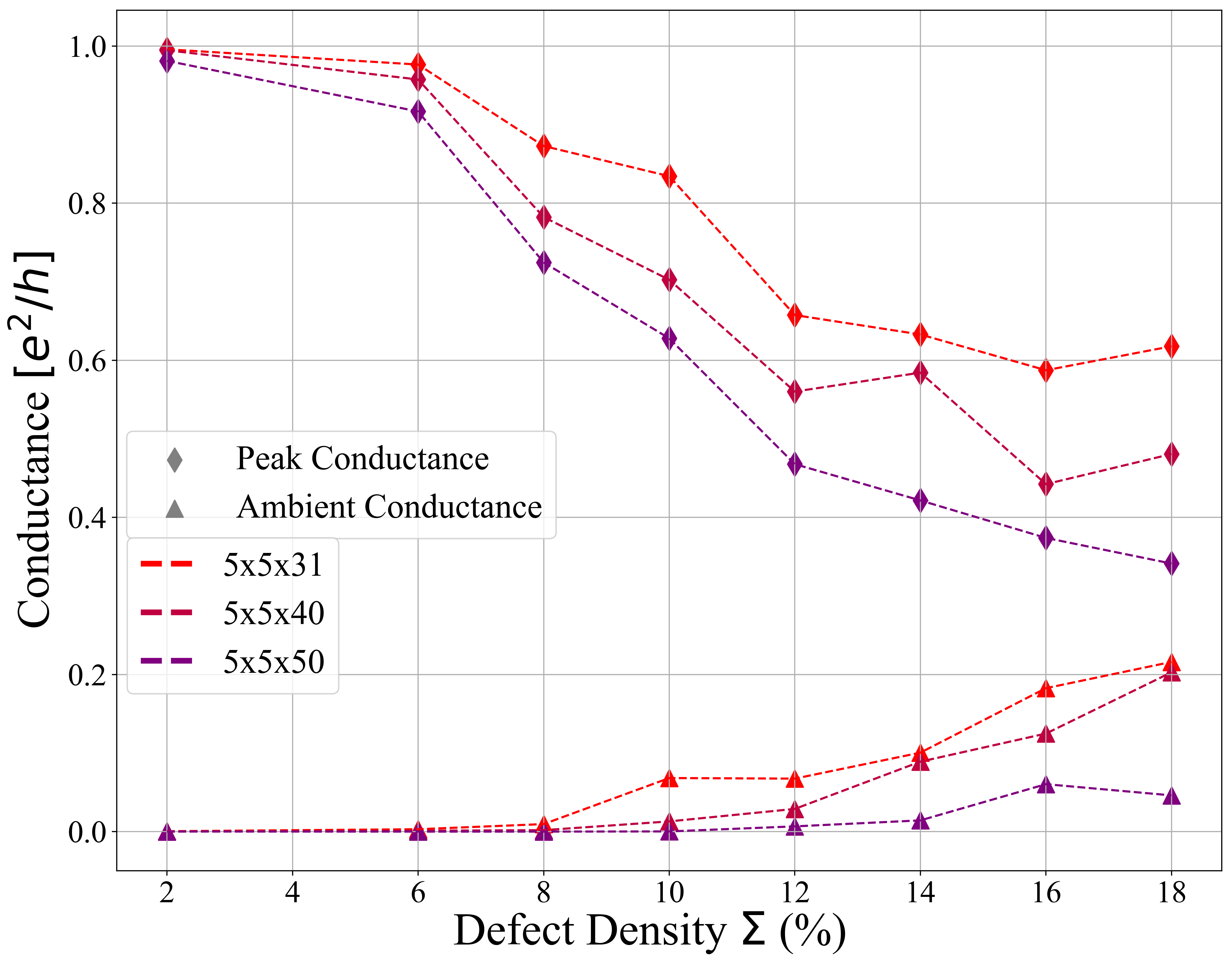}
    \caption{Ensemble averaged minimum (ambient) and maximum (peak) conductance values for three wire sizes which support the expected trend of decreasing peak conductance values with increasing connectivity defect density accompanied by an increase in ambient conductance.}
    \label{fig:anal}
\end{figure}

It is noticed while inspecting signals such as those shown in Figure \ref{fig:trans_single}, that the while the signal has the same periodicity, the peak conductance is lower for higher connectivity defect densities once we have taken a temperature average. To characterize this feature, we compute the high temperature conductance signal for a variety of wires and vary the connectivity defect density and obtain Figure \ref{fig:trans}. The signals in Figure \ref{fig:trans} can be interpreted as the expected conductance signature for an ensemble of wires at high temperature. We observe that the signals have the same periodicity as the expected ideal signals for a crystalline system. Further, there is a statistically expected decrease in peak conductance with increasing temperature and also with increasing wire length. We also observe that ambient (minimum) conductance increases with increasing connectivity defect density. Thus, we are led to conclude that increasing the connectivity defect density leads to a \textit{flattening} of the expected signal.

We collect the ambient and peak values for conductance signatures of various wires in the range $[0,2\Phi_0]$ and confirm that the expected trend is realised in Figure \ref{fig:anal}.

\section{Discussion}
We have developed evidence that coherent Aharonov-Bohm oscillations in the conductance of a 3D topological insulator wire persist with the introduction of connectivity defects using a lattice model with fixed coordination number. Further we have demonstrated that the measured conductance displays a flattening behaviour, as described in the previous section. 

We expect that the conclusions of this paper hold in other amorphous models in the same topological phase as this model. For instance, another amorphous model of interest was first described in \cite{agarwala2017} and then slightly modified in \cite{hannukainen2022local}. This model is fully `amorphous' but does not have a fixed coordination number. It would be of interest for future work to characterize the transport behaviour of this model to evaluate the effect of the specific implementation of amorphous disorder.

Further the trends relating the connectivity defect density and the flattening of the signal are only demonstrated here numerically and it would be fruitful to have a first principles explanation for this expectation. We hope that this work prompts experiments towards observing and characterizing this effect leading to further exploration into the features and distinctions of amorphous topological materials.

We thank Anton Akhmerov, H\'{e}l\`{e}ne Spring, Isidora Araya, Sin\'{e}ad Griffin, Rahul Sahay, Iryan Na, Omar Ashour, Adolfo Grushin and Alexander Avdoshkin for insightful discussions. This work was supported by the U.S. Department of Energy, Office of Science, National Quantum Information Sciences Research Centers, Quantum Science Center. E.J.D was supported by the NSF Graduate Research Fellowship Program, NSF Grant No. DGE 1752814

\bibliographystyle{unsrt}
\bibliography{PRB_template}

\newpage
\onecolumngrid

\end{document}